\documentclass[10pt]{iopart}
\usepackage{graphicx}

\begin{document}

\title[Stability of ferromagnetism in the half-metallic pnictides]
{Stability of ferromagnetism in the half-metallic pnictides and
similar compounds: A first-principles study}

\author{E \c Sa\c s\i o\~glu\dag,
I Galanakis\ddag, L M Sandratskii\dag\ and P Bruno\dag}

\address{\dag\ Max-Planck Institut f\"ur Mikrostrukturphysik,
D-06120 Halle, Germany}
\address{\ddag\ Institut of Microelectronics, NCSR ``Demokritos'',
15310 Aghia Paraskevi, Athens, Greece}

\ead{ersoy@mpi-halle.de, i.galanakis@fz-juelich.de,
lsandr@mpi-halle.de, bruno@mpi-halle.de}

\begin{abstract}
Based on  first-principles electron structure calculations and
employing the frozen-magnon approximation we study the exchange
interactions in a series of transition-metal binary alloys
crystallizing in the zinc-blende structure and calculate the Curie
temperature within both the mean-field approximation (MFA) and
random-phase approximation (RPA). We study two Cr compounds, CrAs
and CrSe, and four Mn compounds: MnSi, MnGe, MnAs and MnC. MnC,
MnSi and MnGe are isovalent to CrAs and MnAs is isoelectronic with
CrSe. Ferromagnetism is particular stable for CrAs, MnSi and MnGe:
All three compounds show Curie temperatures around 1000 K. On the
other hand,  CrSe and MnAs show a tendency to antiferromagnetism
when compressing the lattice. In MnC the half-metallic gap is
located in the majority-spin channel contrary to the other five
compounds. The large half-metallic gaps, very high Curie
temperatures, the stability of the ferromagnetism with respect to
the variation of the lattice parameter and a coherent growth on
semiconductors make MnSi and CrAs most promising candidates for
the use in spintronics devises.

\end{abstract}




\section{Introduction}\label{sec1}

The emerging field of spintronics \cite{Prinz,deBoeck}, also known
as magneto- or spin-electronics, has provided a strong motivation
towards the fabrication of novel magnetic materials. A strong
interest has been focused on the so-called half-metallic
ferromagnets, \textit{i.e.} metallic compounds for which either
the majority or minority spin-channel presents a gap at the Fermi
level $E_F$ \cite{deGroot}. Such behavior is detected in various
perovskite structures \cite{Soulen}, some oxides like CrO$_2$
\cite{Korotin} and magnetite (Fe$_3$O$_4$) \cite{Yanase}, several
Heusler alloys \cite{GalanakisHalf,GalanakisFull,GalanakisQuat},
or some diluted magnetic semiconductors (DMS) \cite{Akai,SandrBruno}.

In 2000, Akinaga and collaborators succeeded to grow a thin film
of the binary alloy CrAs on top of GaAs \cite{Akinaga}. It was
found that CrAs was actually continuing the zinc-blende (zb)
structure of GaAs crystallizing in a metastable phase. The most
astonishing property was the total spin magnetic moment of 3
$\mu_B$, which is consistent with the half-metallic character
predicted by relevant electronic structure calculations published
in the same paper and reproduced with larger accuracy in latter
publications \cite{Shirai,Shirai2}. The thin films of CrAs have
also the great advantage of a high Curie temperature
$T_\mathrm{C}$, exceeding the 400~K (limit of the magnetometer
used in the experiment) \cite{Mizuguchi}. This discovery intrigued
the interest of experimentalists and several attempts were made to
grow similar metallic compounds as thin films. Zhao \textit{et
al.} have grown zb CrSb films by molecular beam epitaxy and found
a $T_\mathrm{C}$ similar to the CrAs case \cite{Zhao}. Moreover,
the growth of small zb MnAs islands on GaAs substrates \cite{Ono}
has been achieved and CrAs/GaAs multilayers have been fabricated
\cite{Mizuguchi2} showing that coherent heterostructures are
possible. These half-metallic materials combine high values of
$T_\mathrm{C}$ with the coherent growth on top of semiconductors
and as the know-how on the growth increases, they can be
considered as very promising candidates for spintronics
applications.

Several theoretical studies have been devoted to the zinc-blende
compounds of transition elements with group-V and VI elements
(respectively, pnictides and chalcogenides). The calculations of
the total energy and structural optimization show, in agreement
with experiment, that the ground state crystal structure of these
compounds is not of the zinc-blende type. In most cases the
equilibrium crystal structure is of the NiAs type \cite{Landolt}.
This structure does not lead to half-metallicity
\cite{Wei,Sanvito,Continenza,Zhao2,Xie}. The zinc-blende structure
can only be obtained by the epitaxial growth of few layers  on
semiconductors. Galanakis and Mavropoulos studied in detail the
magnetic properties and the position of the Fermi level in zb V-,
Cr- and Mn-pnictides and chalcogenides as a function of the
lattice constant \cite{Galanakis}. Several authors have confirmed
these results \cite{Liu,Pask,Sanyal,Xu,Xie2,Yao,Zhang}. Some of
the binary semiconductors crystallize in the wurtzite structure
and a number of studies is devoted to such systems
\cite{Xie3,Zhang2}. Of special interest is the case of
half-metallic zb MnC compound whose properties differ from the
properties of other Mn compounds \cite{Qian}. Also the cases of
unsupported clusters \cite{Nakao} and nanocrystallites
\cite{Qian2} have been studied.

An important aspect of the studies is the stability of the
half-metallic gap. It has been shown that the tetragonal
deformation of the zb lattice does not destroy the gap
\cite{Shi,Yamana} and  for some systems it even makes the zb
structure more stable \cite{Zhao3,Zheng}. Contrary to other
half-metallic systems, in the case of the zb transition metal
compounds it is possible to retain the half-metallicity both at
the surfaces of these materials \cite{Galanakis,Galanakis2} and in
the case of multilayers with binary semiconductors
\cite{Fong,Mavropoulos,Nagao,Geshi2,Qian3,Bengone}. The V and Cr
atoms have the ability to increase considerably their spin moment
as a result of decreasing number of neighbors keeping in this way
the half-metallicity of the system.

Contrary to Cr and V, the Mn spin moment is saturated in these
compounds that leads to the loss of the half-metallicity at the
surface \cite{Galanakis}. The spin-orbit coupling has very weak
effect on the spin-polarization of the electron states at the
Fermi level for the $sp$ atoms up to 5th period of the Mendeleev's
table \cite{Mavropoulos2}. In the case of MnBi, where Bi is a
heavy element from the 6th period, the spin-orbit coupling leads
to the decrease of the spin-polarization from 100\% to 77\%
\cite{Mavropoulos2}. Recently, Chioncel and collaborators studied
the influence of the correlation effects on the electron structure
of CrAs \cite{Chioncel}. They found that the spin-magnon
interaction leads to the appearance of nonquasiparticle states in
the spin-minority channel. The states are shown to lie above the
Fermi level and to be sensitive to the lattice constant.

In contrast to the gap properties, the exchange interactions in
these compounds attracted much less attention. Shirai has shown
that in the case of pnictides (VAs, CrAs and MnAs) the
ferromagnetic state is energetically preferable compared to the
antiferromagnetic state \cite{Shirai}. Only in the case of zb FeAs
an opposite trend was found \cite{Shirai}. Sakuma demonstrated
that the Mn compounds isoelectronic to CrAs (MnSi, MnGe and MnSn)
are half-metallic ferromagnets \cite{Sakuma}. In these compounds,
similar to CrAs, the long-range ferromagnetism is stabilized by
the short-range direct Mn-Mn interaction and an indirect
interaction through the $sp$ atoms. The Curie temperature in both
CrAs and the isoelectronic MnZ compounds was found to exceed
1000~K \cite{Sakuma}. K\"ubler calculated $T_\mathrm{C}$ for VAs,
CrAs and MnAs and found the largest value of 1041~K in the case of
CrAs compared with 764~K in VAs and 671~K in MnAs \cite{Kubler}.
Taking MnAs as an example, it was shown that the loss of the
half-metallicity in a compressed lattice leads to a strong
decrease of the Curie temperature \cite{Kubler}. Sanyal and
collaborators have calculated $T_\mathrm{C}$ for V-, Cr- and MnAs
compounds assuming the GaAs and InAs lattice constants
\cite{Sanyal}. Their mean field approximation (MFA) results agree
with the calculations of other authors while the values of
$T_\mathrm{C}$ obtained with Monte Carlo simulations are $\sim$100
K lower than the MFA values for VAs and MnAs and $\sim$300 K for
CrAs. Note that the mean-field formula for the Curie temperature
used in references \cite{Sanyal,Sakuma} is derived for a
single-sublattice system. Therefore the exchange interaction
between 3$d$ and $sp$ atoms is not treated consequently. Also the
formula of K\"ubler \cite{Kubler} is derived for an one-sublattice
case. In general, the treatment of a multiple-sublattice materials
needs solving a system of equations with the number of equations
equal to the number of sublattices (section \ref{sec2}). Below we
show that for materials studied in this paper the one-sublattice
approximation is well founded. In other systems, however, it leads
to a large error \cite{Sasioglou2004,Sasioglou2005_1}.

In this work we report the theoretical investigation of the
magnetic exchange interactions for selected zb compounds. We study
two Cr compounds, CrAs and CrSe, and four Mn compounds, MnAs,
MnSi, MnGe and MnC. MnAs is isoelectronic to CrSe whereas other
three Mn compounds are isoelectronic to CrAs. We use the
calculated exchange parameters to estimate the Curie temperature
in both the mean field approximation ($T_\mathrm{C}^\mathrm{MFA}$)
and random phase approximation ($T_\mathrm{C}^\mathrm{RPA}$). A
special attention is paid to the influence of the position of the
Fermi level on the magnetic properties.

The paper is organized as follows: In section~\ref{sec2}, we
discuss the calculational technique. In sections~\ref{sec3} and
\ref{sec4}, we present the results for the Cr and  Mn based
half-metallic compounds. Section \ref{sec5} is devoted to MnC.
Finally we summarize in section~\ref{sec6}.

\section{Calculational Method}\label{sec2}

The calculations are carried out using the augmented spherical
waves method (ASW) \cite{asw} within the atomic--sphere
approximation (ASA) \cite{asa}. The exchange--correlation
potential is chosen in the generalized gradient approximation
\cite{gga}. A dense Brillouin zone (BZ) sampling
$30\times30\times30$ is used. The radii of all atomic spheres are
chosen equal.

The method of the calculation of the exchange constants has been already
presented elsewhere \cite{Sasioglou2004}. Here, to make the paper
reasonably self-contained a brief overview is given.
We describe the interatomic exchange interactions in terms of
the classical Heisenberg Hamiltonian
\begin{equation}
\label{eq:hamiltonian2} H_{eff}=-  \sum_{\mu,\nu}\sum_{\begin
{array}{c}
^{{\bf R},{\bf R'}}\\ ^{(\mu{\bf R} \ne \nu{\bf R'})}\\
\end{array}} J_{{\bf R}{\bf R'}}^{\mu\nu}
{\bf s}_{\bf R}^{\mu}{\bf s}_{\bf R'}^{\nu}
\end{equation}
In equation \ref{eq:hamiltonian2}, the  indices  $\mu$ and $\nu$
number different sublattices and ${\bf R}$ and ${\bf R'}$ are the
lattice vectors specifying the atoms within sublattices, ${\bf
s}_{\bf R}^\mu$ is the unit vector pointing in the direction of
the magnetic moment at site $(\mu,{\bf R})$.

We employ the frozen--magnon approach \cite{magnon} to calculate
interatomic Heisenberg exchange parameters. The calculations
involve few steps. In the first step, the exchange parameters
between the atoms of a given sublattice $\mu$ are computed. The
calculation is based on the evaluation of the energy of the
frozen--magnon configurations defined by the following atomic
polar and azimuthal angles
\begin{equation}
\theta_{\bf R}^{\mu}=\theta, \:\: \phi_{\bf R}^{\mu}={\bf q \cdot
R}+\phi^{\mu}. \label{eq_magnon}
\end{equation}
The constant phase $\phi^{\mu}$ is always chosen equal to zero.
The magnetic moments of all other sublattices are kept parallel to
the z axis. Within the Heisenberg model~(\ref{eq:hamiltonian2})
the energy of such configuration takes the form
\begin{equation}
\label{eq:e_of_q} E^{\mu\mu}(\theta,{\bf
q})=E_0^{\mu\mu}(\theta)+\sin^{2}\theta J^{\mu\mu}({\bf q})
\end{equation}
where $E_0^{\mu\mu}$ does not depend on {\bf q} and the Fourier
transform $J^{\mu\nu}({\bf q})$ is defined by
\begin{equation}
\label{eq:J_q} J^{\mu\nu}({\bf q})=\sum_{\bf R} J_{0{\bf
R}}^{\mu\nu}\:\exp(i{\bf q\cdot R}).
\end{equation}

In the case of $\nu=\mu$ the sum in equation (\ref{eq:J_q}) does
not include ${\bf R}=0$. Calculating $ E^{\mu\mu}(\theta,{\bf q})$
for a regular ${\bf q}$--mesh in the Brillouin zone of the crystal
and performing back Fourier transformation one gets exchange
parameters $J_{0{\bf R}}^{\mu\mu}$ for sublattice $\mu$. The
determination of the exchange interactions between the atoms of
two different sublattices $\mu$ and $\nu$  is discussed in
reference \cite{Sasioglou2004}.

First, the Curie  temperature is estimated within the mean--field
approximation (MFA) for the case of a multi--sublattice material by
solving the system of  coupled   equations \cite{Sasioglou2004,Anderson}
\begin{equation}
\label{eq_system} \langle s^{\mu}\rangle
=\frac{2}{3k_BT}\sum_{\nu}J_0^{\mu\nu}\langle s^{\nu}\rangle
\end{equation}
where  $\langle s^{\nu}\rangle$ is the average $z$ component of
${\bf s}_{{\bf R}}^{\nu}$  and $J_0^{\mu\nu}\equiv\sum_{\bf R}
J_{0{\bf R}}^{\mu\nu}$. Equation \ref{eq_system} can be
represented in the form of eigenvalue matrix problem
\begin{equation}
\label{eq_eigenvalue} ({\bf \Theta}-T {\bf I}){\bf S}=0
\end{equation}
where $\Theta_{\mu\nu}=\frac{2}{3k_B}J_0^{\mu\nu}$, ${\bf I}$ is a
unit matrix and ${\bf S}$ is the vector of $\langle s^{\nu}\rangle
$. The largest eigenvalue of matrix $\Theta$ gives the value of
$T_\mathrm{C}^\mathrm{MFA}$ \cite{Anderson}.

Within the random phase approximation (RPA) the Curie temperature is
given by the relation \cite{pajda}
\begin{equation}
\label{eq_RPA} \frac{1}{k_\mathrm{B}T_\mathrm{C}^\mathrm{RPA}}=
\frac{6\mu_\mathrm{B}}{M}\frac{1}{N}\sum_q\frac{1}{\omega(\mathbf{q})},
\end{equation}
\noindent where $\omega(\mathbf{q})$ is the spin-wave dispersion.
Equation (\ref{eq_RPA}) does not take into account the presence of
several sublattices and will be used to estimate the Curie
temperature with account for the interactions within the
sublattice of the 3$d$ atoms. As we will show these interactions
give a leading contribution to the Curie temperature.

\begin{table}
\caption{Calculated atom-resolved and total spin moments in
$\mu_\mathrm{B}$ for CrAs and CrSe (2$^\mathrm{nd}$ to
6$^\mathrm{th}$ column).  CrAs and CrSe are half-metallic for the
GaAs and CdS experimental lattice constants, respectively.
a$_{II}$ means that the Fermi level is at the upper edge of the
gap and a$_{III}$ corresponds to 1\% contraction with respect to
a$_{II}$. } \label{table1}
\begin{indented}
 \item[]
 \begin{tabular}{ll|ccccc}\br
    Compound     &   a(\AA )  &  $m^\mathrm{Cr}$ & $m^\mathrm{As,Se}$ &
    $m^\mathrm{Void1}$ & $m^\mathrm{Void2}$  &  $m^\mathrm{Total}$  \\ \mr
        CrAs -- a$_{I[GaAs]}$ &  5.65   &  3.24     &   -0.31     &     -0.03
        & 0.10   &  3.00  \\
        CrAs -- a$_{II}$   &   5.53  &   3.17   &     -0.25 &   -0.03 &
0.11  &   3.00  \\
       CrAs -- a$_{III}$  &  5.47   &  3.10  &   -0.23    &   -0.03
& 0.10  &   2.95 \\ \mr
       CrSe -- a$_{I[CdS]}$   & 5.82  &   3.92    &   -0.10  &     0.02 &
0.16  &   4.00  \\
      CrSe --  a$_{II}$    &  5.62  &   3.84   &     -0.04  &     0.02 &
0.18  &   4.00 \\
      CrSe -- a$_{III}$  &     5.56  &   3.77   &     -0.04 &     0.02 &
0.16  &   3.92  \\ \br
\end{tabular}
\end{indented}
\end{table}

\section{Cr-compounds: CrAs and CrSe} \label{sec3}

We begin with the discussion of the Cr compounds: pnictide CrAs
and chalcogenide CrSe. The half-metallicity in these compounds
results from the formation of bonding and antibonding hybrides
between the $t_{2g}$ states ($d_{xy}$, $d_{yz}$, and $d_{xz}$) of
the transition metal atom (here Cr) and the  $p$ states of the
four neighboring $sp$ atoms (As or Se). The hybridizational gap is
partly filled by the $e_g$ states of the $3d$ atom ($d_{z^2}$ and
$d_{x^2-y^2}$). The hybridization takes place in both spin
channels. The position of the hybridization gap is different for
majority and minority electrons as a consequence of the exchange
splitting \cite{Galanakis}. There are in total four occupied
minority states, one coming from the $s$ states of the $sp$ atom
and three bonding 3\textit{d} states. The minority-spin $e_g$
orbitals lay above the Fermi level. Thus the total spin moment
$M_t$ in $\mu_\mathrm{B}$ is given by the relation
$M_t=(Z_{\mathrm{tot}}-8)\ \mu_B$, where $Z_{\mathrm{tot}}$ is the
total number of valence electrons in the unit cell
\cite{Galanakis}. In the cases of CrAs (11 valence electrons) and
CrSe (12 valence electrons) the total spin moment should be 3 and
4 $\mu_B$, respectively.

\begin{figure}
  \begin{center}
    \includegraphics[scale=0.36]{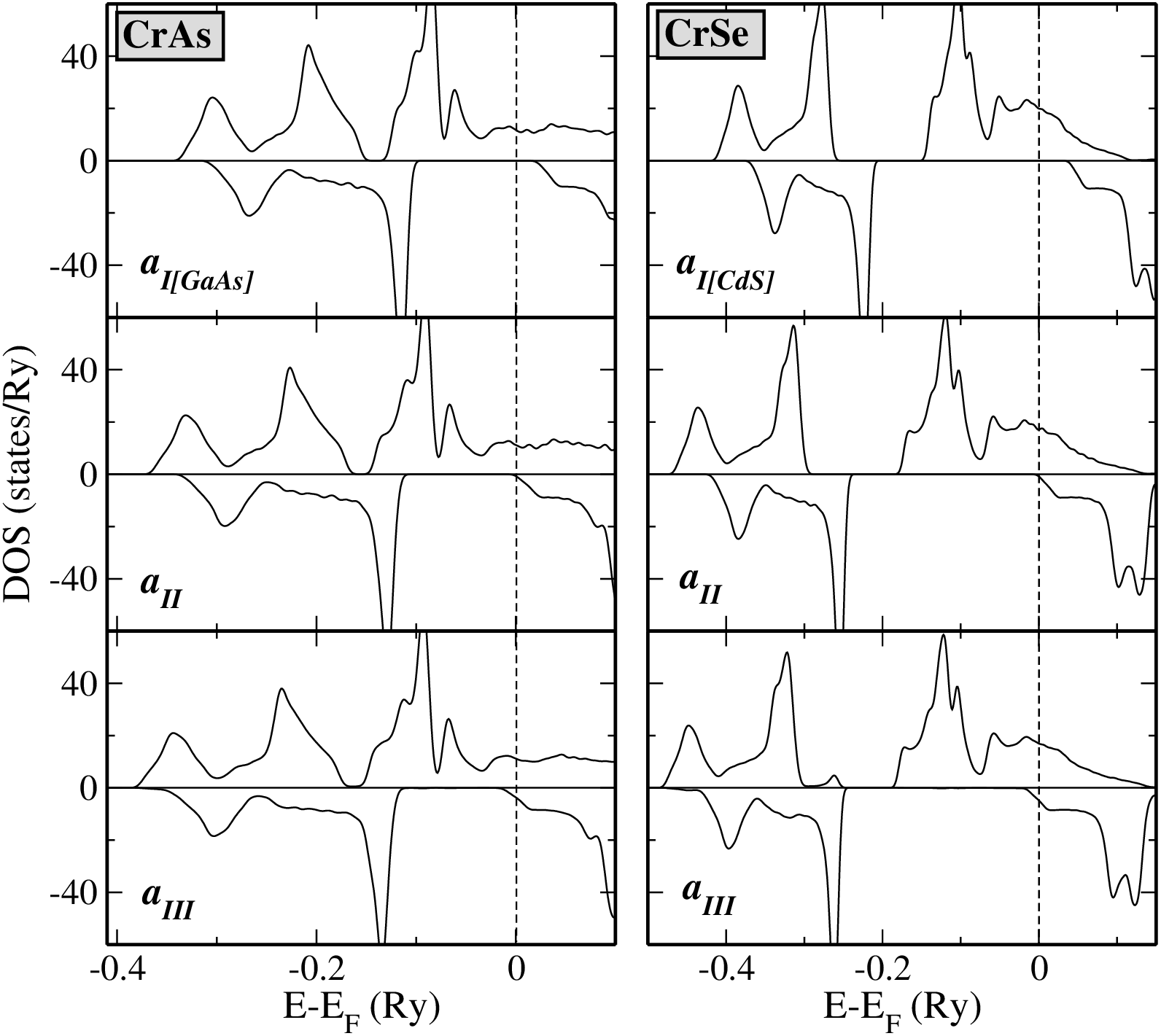}
  \end{center}
\caption{Calculated spin-resolved density of states of CrAs and
CrSe. The upper panel corresponds to the experimental lattice
constant of GaAs and CdS, respectively, the middle panel to the
a$_{II}$ case in table \ref{table1} (Fermi level at the
higher-energy edge of the half-metallic gap) and the bottom to the
a$_{III}$ case in the same table (1\% contraction of the lattice
constant with respect to the a$_{II}$ case).} \label{fig1}
\end{figure}

In table \ref{table1} we present the atom-resolved and total spin
moments. Notations Void1 and Void2 are used for two types of empty
spheres used in our ASA calculations to describe correctly the
charge distribution in the zinc-blende structure. For each system
we performed calculations for three different lattice parameters.
The first parameter (case I) corresponds to the experimental
lattice constant of GaAs(CdS) for CrAs(CrSe). For these lattice
constants both compounds are half-metals. This is seen in figure
\ref{fig1} where the spin-projected density of states (DOS) is
presented.  The total spin moment is exactly integer as expected
for a half-metallic system. The Cr atoms posses a very large spin
moment of 3.24 $\mu_\mathrm{B}$ in CrAs that reaches the value of
3.92 $\mu_\mathrm{B}$ in CrSe. A larger moment of CrSe is a
consequence of one extra valence electron per formula unit. Both
As and Se atoms have small induced moments antiparallel to the Cr
moments. The discussion of the spin moments of the $sp$ atoms and
their relative magnitude can be found in reference
\cite{Galanakis}.

The half-metallic gap of CrSe is much larger than for CrAs. This
is explained by the larger Cr moment in CrSe that leads to a
larger exchange splitting $\Delta E_x$ and, as a result, to a
higher position of the minority $e_g$ bands.

In the next step, we compress the lattice simulating a hydrostatic
pressure. The contraction of the lattice constant results in a
higher energy position of the majority $p$ states with a
corresponding increase of the Fermi energy, $E_\mathrm{F}$. The
calculations are performed for two decreased values of the lattice
parameter. The first (case II) corresponds to the position of the
Fermi level at the bottom edge of the minority-spin conduction
band (fig. \ref{fig1}). The second parameter (case III) is
obtained by a further 1\% decrease of the lattice constant. In the
latter case the Fermi level lies within the spin-minority
conduction band. Since half-metallicity is lost the total magnetic
moment in case III becomes non-integer (table \ref{table1}).

The contraction of the lattice leads to an increase of the
hybridization between the Cr states and the states of the $sp$
atoms. As the result, both the Cr moment and the induced moment of
the $sp$ atom decrease. In case II, the changes of different
moments compensate resulting in the same integer total moment. The
decrease of the Cr spin moment causes a small decrease of the
width of the gap.

\begin{figure}
  \begin{center}
    \includegraphics[scale=0.4]{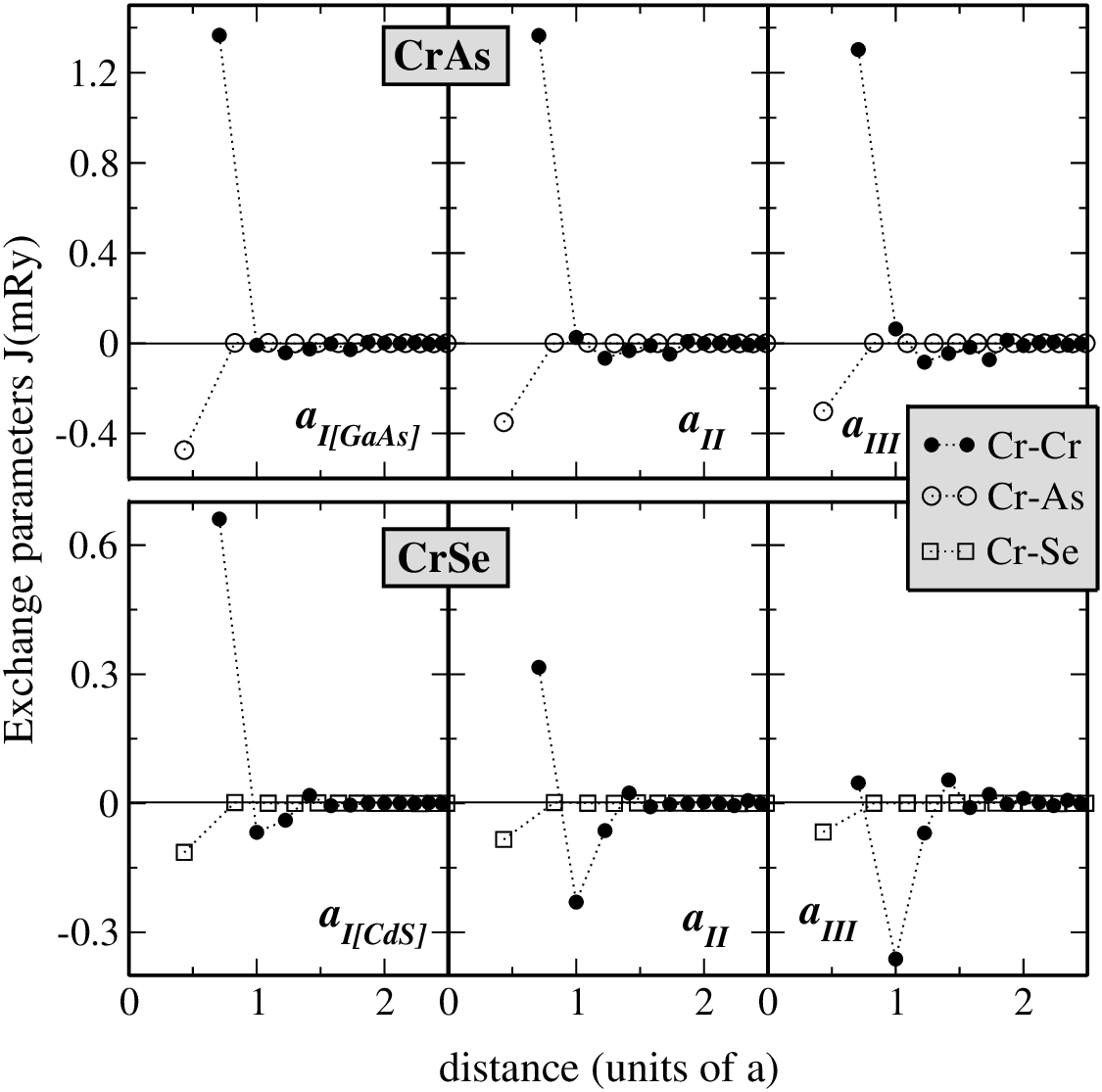}
  \end{center}
\caption{The exchange constants of CrAs (upper panel) and CrSe
(lower panels) for three different lattice constants. The left
panel corresponds to the experimental lattice constant of GaAs and
CdS respectively for CrAs and  CrSe, the middle one to
case II and the right one to case III.} \label{fig2}
\end{figure}

Next we used the frozen-magnon approach to calculate the exchange
parameters for both systems and three values of the lattice
constants (figure \ref{fig2}). The exchange interaction between
the induced moments of the $sp$ atoms is very small and the
corresponding parameters are not presented.

For the semiconductor lattice parameter (left panels in figure
\ref{fig2})  the interactions are characterized by the large
ferromagnetic coupling of the nearest Cr atoms. In CrAs, this
coupling is about twice the coupling in CrSe. In the zinc-blende
structure the nearest surrounding of a Cr atom is formed by 4 $sp$
atoms. The second-nearest coordination sphere contains Cr atoms.
Although the exchange interaction Cr-$sp$ atom is
antiferromagnetic it makes the parallel directions of the Cr
moments energetically preferable. Therefore the indirect exchange
interaction of the Cr moments through the $sp$ atoms stabilizes
further the ferromagnetic configuration of the Cr moments (a
similar situation was found in full-Heusler alloys
\cite{Kubler2}). The interaction is short range and involves only
the nearest Cr atoms.

The influence of the lattice compression is very different for the
two systems. In CrAs, the compression leads to a small decrease of
the exchange constants revealing the robustness of the
ferromagnetism of the compound. On the other hand, in CrSe the
effect of the compression is very strong. Already for the a$_{II}$
case where the system is still half-metallic the exchange
interaction between the nearest Cr atoms is halved with respect to
the case of the CdS lattice constant. The interaction between
second-nearest Cr atoms strongly increases keeping
antiferromagnetic character. This is reflected in a very small
calculated Curie temperature (table \ref{table2}). When the
lattice is further compressed and the Fermi level moves into the
minority-spin conduction band, the ferromagnetic interaction
between the nearest Cr atoms almost vanishes whereas the
antiferromagnetic interaction between the second nearest Cr atoms
increases further. This leads to the instability of the
ferromagnetic state.

The calculated Curie temperatures are collected in table
\ref{table2}. They reflect the properties of the exchange
interactions discussed above. We report two estimations of the
Curie temperature within the MFA, one with only the Cr-Cr
interactions taken into account and other with both Cr-Cr and
Cr-$sp$ atom interactions considered, and one RPA estimation.

In case I, for both systems the Curie temperature is much higher
than the room temperature. For CrAs,  $T_\mathrm{C}$ is about
twice larger than in CrSe and exceeds 1000 K. The compression of
the lattice leads in the case of CrAs to a very moderate decrease
of the Curie temperature. In CrSe, the $T_\mathrm{C}$ drops
strongly with decreasing lattice parameter becoming negative in
case III revealing the instability of ferromagnetism.

Two mean-field estimations give very similar results (table
\ref{table2}) revealing the leading role of the Cr-Cr
interactions. Therefore the neglect of the inter-sublattice
Cr-$sp$ atom  interactions does not result in a substantial error
for these compounds.

\begin{table}
\caption{Calculated Curie temperatures for CrAs and CrSe. In the
3$^\mathrm{rd}$ and 4$^\mathrm{th}$ columns the MFA-estimations of
$T_\mathrm{C}^\mathrm{MFA}$ are presented (respectively, with only
the Cr-Cr interactions and with both the Cr-Cr and Cr-As(Se)
interactions taken into account. In the 5$^\mathrm{th}$ column,
the RPA-estimation of $T_\mathrm{C}$ is presented. Negative values
of $T_\mathrm{C}$ reveal the instability of the ferromagnetic
structure.} \label{table2}
\begin{indented}
\item[]
 \begin{tabular}{ll|ccc} \br
 Compound       &   a(\AA )  &
    $T_\mathrm{C}^\mathrm{MFA}$(Cr-Cr)&
$T_\mathrm{C}^\mathrm{MFA}$ & $T_\mathrm{C}^\mathrm{RPA}$ \\ \mr
CrAs --  a$_{I[GaAs]}$  &  5.65   & 1576&  1600&   1176 \\
CrAs -- a$_{II}$   &   5.53  & 1498   & 1512&   1011 \\
CrAs --  a$_{III}$  &  5.47   &1376 & 1387& 860  \\ \mr
CrSe --  a$_{I[CdS]}$   & 5.82  &  706 & 709&  508 \\
CrSe -- a$_{II}$    &  5.62  &   109 & 117 & 24\\
CrSe --  a$_{III}$  &     5.56    & -248&  -251&-194 \\ \br
\end{tabular}
\end{indented}
\end{table}

Since the intra-sublattice interactions dominate, the use of the
RPA formula (\ref{eq_RPA}) is justified. Overall the
$T_\mathrm{C}$ values within RPA are by 20-30\% smaller than the
MFA ones.  The difference between MFA and RPA estimations is
explained by a different weighting of the spin-wave excitations.
In the MFA all excitations are taken with equal weight whereas in
the RPA the weight decreases with increasing energy of the
excitation \cite{pajda,bouzerar}.

The RPA is expected to provide better estimation of the Curie
temperature. In section \ref{sec5}, we will discuss the relation
between the MFA and RPA values of $T_\mathrm{C}$ in more detail.
The only experimental information about the magnetic transition
temperature of the systems considered is that the Curie
temperature of CrAs is well above 400 K \cite{Akinaga} which is in
good correlation with our estimations.

In the case of CrAs we can compare the calculated $T_\mathrm{C}$
values of 1600 K within MFA and 1176 K within RPA with the results
of previous ab-initio studies. Sanyal and collaborators calculated
the Curie temperature within MFA and found a value of 1320 K for
the GaAs experimental lattice constant and 1100 K for the InAs
experimental lattice constant \cite{Sanyal}. Their Monte Carlo
simulations gave for the same lattice constants 980 K and 790 K,
respectively. K\"ubler used a Ginzburg-Landau approach with a
simplified form of the dynamical susceptibility and a relaxation
parameter fitted to reproduce the experimental Curie temperature
of NiMnSb.  His estimation of $T_\mathrm{C}$ of CrAs gives 1041 K
\cite{Kubler}. Finally, Sakuma's mean-field calculation for the
theoretical equilibrium lattice constant of 5.82 \AA\ gave
$T_\mathrm{C}$ about 1400 K \cite{Sakuma}. Summarizing, all
calculations agree that the Curie temperature of CrAs should
exceed or be about 1000 K. A high $T_\mathrm{C}$ value that is
stable with respect to the variation of the lattice constant makes
CrAs an interesting candidate for spintronics applications.

\begin{table} \caption{Similar to table \ref{table1} for the
Mn-based compounds.} \label{table3}
\begin{indented}
\item[]
 \begin{tabular}{ll|ccccc}\br
  Compound      &   a(\AA )  &  $m^\mathrm{Mn}$ & $m^\mathrm{Z}$ &
    $m^\mathrm{Void1}$ & $m^\mathrm{Void2}$  &  $m^\mathrm{Total}$  \\
    \mr
  MnAs --      a$_{I[InP]}$   &   5.87  &   4.16  &      -0.26   &
        -0.01& 0.10    & 4.00\\
     MnAs --     a$_{II}$     &    5.73 &    4.09   &     -0.20   &     -0.01
& 0.11  &   4.00  \\
 MnAs --         a$_{III}$    &     5.68   &  4.01  &      -0.19  &        -0.01
& 0.10  &   3.91  \\ \mr
       MnGe --   a$_{I[GaAs]}$ &  5.65  &   3.43   &     -0.39   &       -0.08
        & 0.04   &  3.00  \\
MnGe --  a$_{II}$  &    5.61   &  3.40   &     -0.37    &
-0.07
& 0.04  &   3.00 \\
  MnGe --      a$_{III}$   &      5.55 &    3.35  &   -0.34  &     -0.07 &
0.04   &  2.98  \\ \mr
  MnSi --        a$_{I[GaAs]}$  & 5.65  &   3.51   &     -0.49  &    -0.07 &
0.05  &   3.00  \\
   MnSi --       a$_{II}$  &        5.52  &   3.42  &    -0.42 &     -0.06 &
0.06  &   3.00 \\
   MnSi --       a$_{III}$ &      5.46  &   3.37   &     -0.39  &  -0.06 &
0.05  &   2.97  \\  \br
\end{tabular}
\end{indented}
\end{table}

\section{Mn-compounds: MnSi, MnGe and MnAs} \label{sec4}

We proceed with the discussion of Mn compounds. Here we study MnSi
and MnGe which are isoelectronic with CrAs and MnAs which is
isoelectronic with CrSe.

\begin{figure}
  \begin{center}
  \includegraphics[scale=0.44]{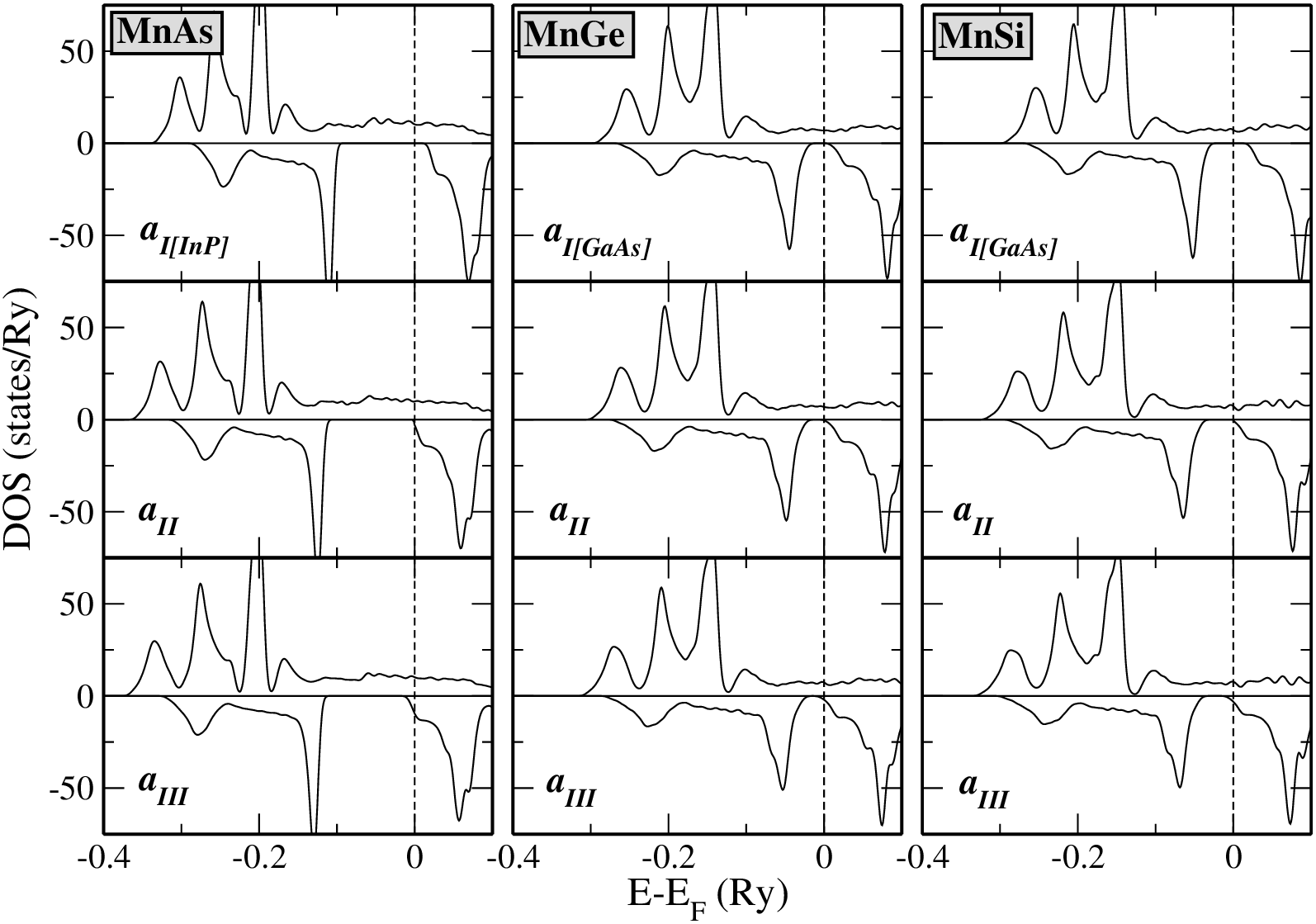}
  \end{center}
\caption{The density of states of MnAs, MnGe and MnSi. See the
caption in figure \ref{fig1} for the description of details.}
\label{fig3}
\end{figure}

In table \ref{table3} we present the atom-resolved and total
magnetic moments. MnAs with the lattice constant of InP, and MnSi
and MnGe with the lattice constant of GaAs are half-metallic and
their total spin moments following the relation presented above
are 4 $\mu_\mathrm{B}$ for MnAs and 3 $\mu_\mathrm{B}$ for MnSi
and MnGe (see table \ref{table3} and figure \ref{fig3}). The Mn
spin moment is about 3.5 $\mu_\mathrm{B}$ for MnSi and MnGe, and
exceeds the 4 $\mu_\mathrm{B}$ in the case of MnAs reflecting the
presence of one more valence electron in the latter compound. The
moments of the $sp$ atoms are antiferromagnetically coupled to the
Mn moments. The value of the half-metallic gap is determined by
the value of the exchange splitting and the energy position of the
$p$ states of the $sp$ atom while the value of the exchange
splitting is governed by the value of the Mn moment. As a result,
the largest gap is obtained for MnAs followed by MnSi and the
smallest gap is obtained for MnGe.

As for the Cr compounds the contraction of the lattice increases
the energy of the majority $p$ states leading to an increased Fermi
level. For all three compounds the hybridization between the $p$
states of the $sp$ atom and the $t_{2g}$ states of Mn increases
and the Mn spin moment slightly decreases with the lattice
contraction. The smaller Mn moments and, as a result, smaller
exchange splittings lead to slightly smaller gap-widths.

\begin{figure}
  \begin{center}
  \includegraphics[width=\textwidth]{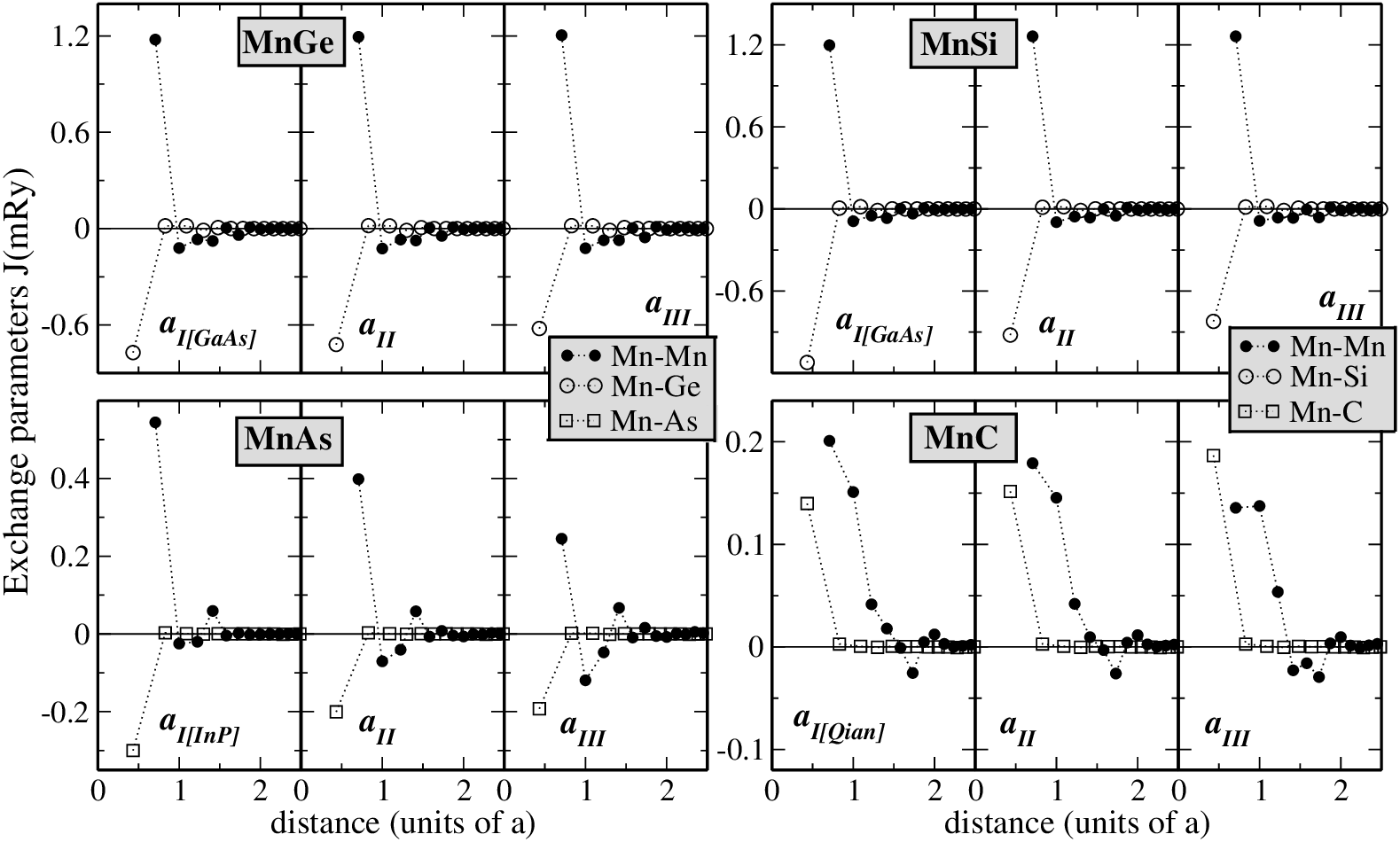}
  \end{center}
\caption{Variation of the exchange constants for the Mn-based
compounds. Left panels correspond to the experimental lattice
constants of semiconductors (see table \ref{table3}), middle panel
to case II and the right panel to case III. } \label{fig4}
\end{figure}

In figure \ref{fig4} we present the calculated exchange constants.
The exchange interaction between the induced moments of $sp$ atoms
is very weak and is not presented. There is strong similarity in
the behavior of the exchange parameters for the Mn compounds
considered here and the Cr compounds discussed in the preceding
section. The properties of MnGe and MnSi are analogous to the
properties of the isovalent CrAs while properties of MnAs are
similar to the properties of isovalent CrSe. Therefore the
corresponding parts of the discussion of the Cr compounds remain
valid also for the Mn systems.

Ferromagnetism is stabilized by both the direct Mn-Mn exchange
interaction and by the indirect interaction through the $sp$
atoms. The exchange interactions are short-range and the Mn-Mn
exchange constants for MnAs are about a factor two smaller than
for MnSi and MnGe.

The compression has little effect on MnSi and MnGe in contrast to
MnAs. Again there is a strong similarity between MnAs and CrSe
although the dependence of the exchange parameters on the lattice
constant in MnAs is somewhat weaker. In particular, the
antiferromagnetic exchange interaction between second-nearest Mn
atoms developing with contraction has lower value than in CrSe.
This results in preserving the ferromagnetic ground state of MnAs
for all lattice parameters considered while in CrSe the
ferromagnetic state becomes unstable (table \ref{table2}).

The Curie temperatures of the Mn compounds are collected in table
\ref{table4}. Again, as for Cr compounds, the Mn-Mn interaction is
dominant and the account for the exchange interaction Mn-$sp$ atom
influences the value of the Curie temperature weakly. In agreement
with the properties of the exchange constants the Curie
temperatures of MnGe and MnSi depend weakly on the lattice
parameter in contrast to MnAs where this dependence is strong. The
Curie temperature of MnAs at highest contraction considered
assumes the value of about $\frac{1}{4}$th of the Curie
temperature at the semiconductor lattice constant. The RPA gives
values of $T_\mathrm{C}$ that are about 30 \% smaller than the
corresponding values within the MFA. For MnSi and MnGe, also the
RPA value of $T_\mathrm{C}$ remains considerably above the room
temperature at all lattice constants studied. The stability of the
high $T_\mathrm{C}$ value with respect to the variation of the
lattice constant makes MnGe and MnSi promising candidates for
spintronics applications. MnSi has an additional advantage of a
larger half-metallic gap.

Next we compare the results of our calculation of $T_\mathrm{C}$
with the results of previous calculations where available. For
MnAs,  we can compare our MFA-estimation of  701 K and RPA
estimation of 551 K obtained for the InP lattice constant with the
Curie temperatures reported by Sanyal et al \cite{Sanyal}. The
MFA-estimation of  Sanyal et al gave the value of 640 K for the
InAs lattice parameter which is very close to the parameter used
in our calculations. Their Monte Carlo simulation gave a
$T_\mathrm{C}$ of 530 K which is very close to our RPA estimation.
K\"ubler obtained the value of 671 K for MnAs with the InAs
lattice constant. When he compressed the lattice down to the GaAs
lattice constant of 5.65\AA\ the $T_\mathrm{C}$ dropped sharply to
210 K \cite{Kubler} in agreement with our results. Finally,
Sakuma's MFA-estimation of the Curie temperature of MnGe for a
lattice constant of 5.82 \AA\ gave a value of about 1200 K which
is very close to our MFA value of 1234 K \cite{Sakuma}. Thus the
results of different calculations are in reasonable agreement with
one another.

\begin{table}
\caption{Calculated spin moments for MnAs, MnGe and MnSi. See the
caption in table  \ref{table2} for the description.}
\label{table4}
\begin{indented}
\item[]
\begin{tabular}{ll|ccc}\br
  Compound      &   a(\AA )  &
    $T_\mathrm{C}^\mathrm{MFA}$(Mn-Mn)&
$T_\mathrm{C}^\mathrm{MFA}$ & $T_\mathrm{C}^\mathrm{RPA}$ \\ \mr
   MnAs --      a$_{I[InP]}$   &   5.87  &  679 & 701 &  551\\
      MnAs --   a$_{II}$     &    5.73  & 400 & 417 &  275 \\
      MnAs --  a$_{III}$    &     5.68   & 201& 230 & 136 \\ \mr
  MnGe --      a$_{I[GaAs]}$ &  5.65   & 1160 &1234 & 716 \\
 MnGe --  a$_{II}$  &    5.61    & 1180 & 1241 &  740\\
     MnGe --   a$_{III}$   &      5.55 &  1184 &1231 & 695 \\ \mr
 MnSi --       a$_{I[GaAs]}$  & 5.65  &   1255 & 1402& 848 \\
      MnSi --     a$_{II}$  &        5.52   &  1310&  1401 & 857 \\
 MnSi --          a$_{III}$ &      5.46   & 1291 & 1367 &  911 \\
 \br
\end{tabular}
\end{indented}
\end{table}

\section{MnC}\label{sec5}

\begin{figure}
  \begin{center}
    \includegraphics[scale=0.36]{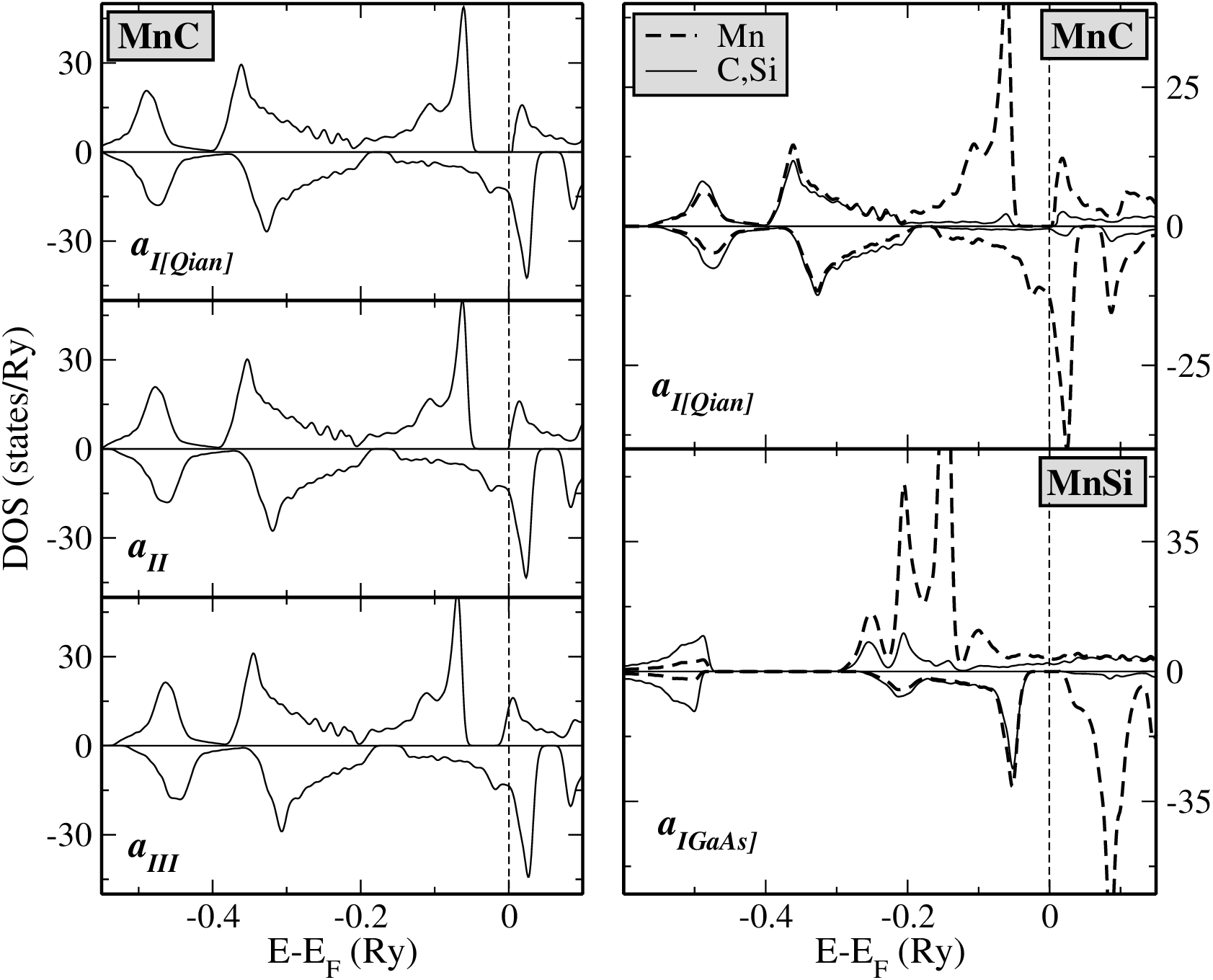}
  \end{center}
\caption{Left panel: calculated spin-resolved DOS of MnC. The
upper panel corresponds to the lattice constant given in reference
\cite{Qian}, the middle and lower panels to the a$_{II}$ and
a$_{III}$ cases in table \ref{table5}.  Right panel: Atom- and
spin-resolved DOS for the half-metallic MnC and MnSi compounds.}
\label{fig5}
\end{figure}

The last Heusler compound studied in this paper is MnC. This
system was predicted to be half-metallic by Qian and collaborators
\cite{Qian}. First we discuss the results of the calculation with
the theoretical equilibrium lattice constant reported by Qian et
al. The magnetic properties of MnC differ from the properties of
the Mn compounds discussed above. MnC has 11 valence electrons: 6
of majority-spin and 5 minority-spin electrons. This electron
distribution results in a magnetic moment which is much smaller
than in the compounds discussed above (table \ref{table5}). In
contrast to the other compounds the half-metallic gap is now in
the majority-spin channel (figure \ref{fig5}). The majority-spin
DOS has the following structure: low in energy in the interval
[-0.55 Ry,-0.4 Ry] lie $s$-states. Above, in the interval [-0.4
Ry,-0.2 Ry] there are three bonding hybrids between the $p$ states
of C and the $t_{2g}$ states of Mn. This is seen in the
atom-projected DOS of MnC (right panel in figure \ref{fig5}) where
both Mn and C states have similar contribution in this energy
interval. The group of states next in energy  (interval [-0.2
Ry,-0.05 Ry]) does not hybridize with the lower states. These are
the Mn $e_g$ states separated by a gap from the antibonding C
$p$-Mn $t_{2g}$ hybrids. The Fermi level lies within this gap in
the  majority-spin DOS. For the minority-spin channel, the
position of the Fermi level belongs to the region of the $e_g$
states. These states are shifted by the exchange splitting to
higher energies with respect to the corresponding majority-spin
states.

The comparison of the DOS of half-metallic MnC and MnSi (figure
\ref{fig5}) allows to reveal the reason for the move of the
semiconductor-type behavior from the spin-minority channel in MnSi
to the spin-majority channel in MnC. First, the two systems differ
by the value of the magnetic moment that leads in the case of MnC
to a smaller relative exchange shift of the spin-up and spin-down
states and to the occupation of a larger number of the spin-down
Mn 3$d$ states. In particular, the minority-spin $e_g$ states are
partly occupied in MnC whereas in MnSi these states lie above the
half-metallic gap. Second, there is difference in the position of
the  $e_g$ states in the hybridizational gap between bonding and
antibonding $t_{2g}$-$p$ hybrids which in the case of MnSi results
in the gap between occupied majority  $e_g$ and empty antibonding
hybrid states.

To extend the study of MnC to the states with the Fermi level
lying at the top-edge of the half-metallic gap (case II) and above
it (case III) we need in this case to expand the lattice (table
\ref{table5}) opposite to all systems considered above where the
lattice must be compressed. This difference comes from the move of
the semiconductor channel to the majority-spin subsystem. The
expansion of the lattice makes the energy bands narrower and
decreases the strength of the $3d$-$p$ hybridization. At the same
time the Mn moment increases increasing the exchange splitting
between majority and minority states. These properties lead to
increasing occupation of the majority-spin states and the loss of
the half-metallicity (figure \ref{fig5}).

The calculated exchange constants are presented in  figure
\ref{fig4}. The ferromagnetism is  stabilized by the direct Mn-Mn
interactions and the Mn-C ferromagnetic coupling. Since the Mn
moment is much smaller than in other compounds the value of the
Mn-Mn exchange interaction is substantially lower (figure
\ref{fig4}). On the other hand, in MnC the first and second
nearest Mn atoms are characterized by a similar strength of the
ferromagnetic exchange interaction while in other systems only the
nearest Mn atoms interact strongly. The expansion of the lattice
leads to decreasing Mn-Mn exchange interactions.

The calculated Curie temperatures are presented in table
\ref{table5}. The Curie temperature of MnC is smaller than in
other Mn compounds being about 500 K (it is comparable with the
RPA estimation of $T_\mathrm{C}$ of MnAs). The value of the Curie
temperature decreases with expansion of the lattice showing again
that half-metallicity is favorable for high Curie temperature.

\begin{table} \caption{Calculated spin magnetic moments and Curie
temperatures for MnC. The value of 4.20 \AA\ is the equilibrium
lattice constant calculated by Qian \textit{et al.} \cite{Qian}.}
\label{table5}
\begin{indented}
\item[]
 \begin{tabular}{ll|ccccc}\br
  Compound      &   a(\AA )  &  $m^\mathrm{Mn}$ & $m^\mathrm{Z}$ &
    $m^\mathrm{Void1}$ & $m^\mathrm{Void2}$  &  $m^\mathrm{Total}$  \\
    \mr
 MnC --  a$_{I(ref. \cite{Qian})}$   &   4.20    & 1.16  & -0.14 & -0.01 & -0.01 &    1.00 \\
MnC --  a$_{II}$   &        4.23  &   1.16   &     -0.14   & -0.01
& -0.01   &  1.00   \\
      MnC --   a$_{III}$   &      4.27   &  1.13   &     -0.16   &       -0.01
& -0.01 &    1.05  \\ \br
 Compound      &   a(\AA )  &
    \multicolumn{3}{r}{$T_\mathrm{C}^\mathrm{MFA}$(Mn-Mn)} &
$T_\mathrm{C}^\mathrm{MFA}$ & $T_\mathrm{C}^\mathrm{RPA}$ \\ \mr
  MnC --   a$_{I(ref. \cite{Qian})}$   &    4.20 &  & &  520&  526 & 507  \\
 MnC --  a$_{II}$   &        4.23   & &&  467 & 476& 458  \\
      MnC --    a$_{III}$   &      4.27  & & & 346&  363 & 327 \\ \br
\end{tabular}
\end{indented}
\end{table}

A remarkable feature of the calculation for MnC is a very small
difference between the $T_\mathrm{C}$ values estimated within the
MFA and the RPA. In section \ref{sec3}  we briefly discussed the
reason behind  the different  values  of  the Curie temperatures
estimated within MFA and RPA. A striking contrast between MnC and
the systems considered in the previous sections allows gaining a
deeper insight into the physics of the formation of the Curie
temperature. We will base our analysis on the comparison of two
systems: MnC and MnSi. Characterizing the relative difference of
the MFA and RPA values of the Curie temperature by the relation $
[T_\mathrm{C}^\mathrm{MFA}-T_\mathrm{C}^\mathrm{RPA}]/T_\mathrm{C}^\mathrm{RPA}$
we get for MnC a small value of 2.5\% compared to a large value of
47\% in MnSi.

In MFA, the Curie temperature is determined by an arithmetic
average of the magnon energies while in RPA  $T_\mathrm{C}$ is
determined by the harmonic average of the same quantities. It is
an arithmetic property that the MFA estimation is larger than the
RPA one or equal to it if all averaged numbers are equal to each
other. In terms of magnon energies, $T_\mathrm{C}^\mathrm{MFA}$ is
equal to $T_\mathrm{C}^\mathrm{RPA}$ in the case that the magnon
spectrum is dispersion-less: the magnon energy does not depend on
the wave vector ${\bf q}$.

\begin{figure}
  \begin{center}
    \includegraphics[scale=0.36]{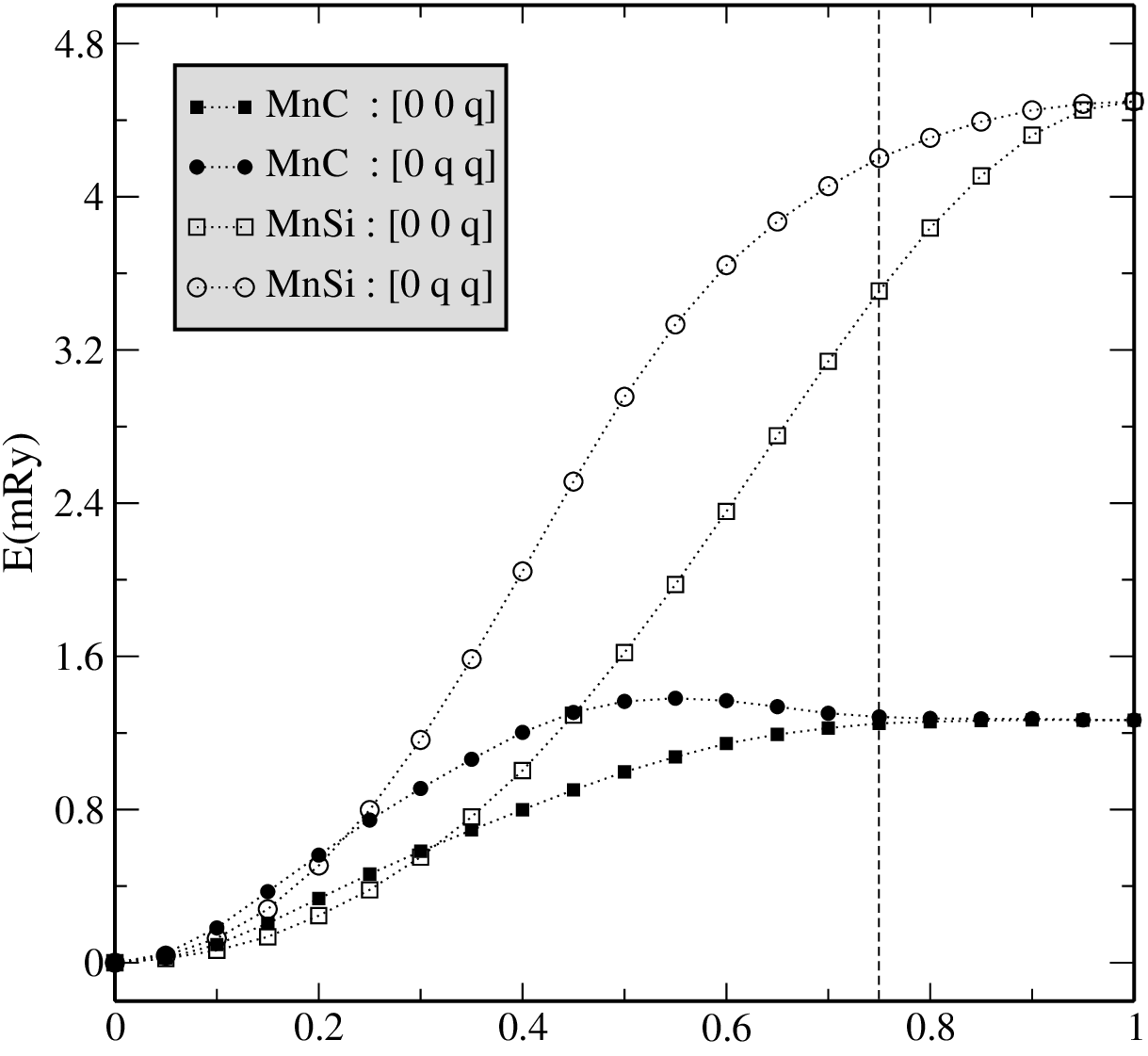}
  \end{center}
\caption{Frozen--magnon  energies   in MnC and MnSi  for [0 0 q]
and [0 q q] directions in the Brillouin zone. The vertical broken
line shows the Brillouin zone boundary for the [0 q q] direction.}
\label{fig6}
\end{figure}

In  figure~\ref{fig6} we plot the  frozen--magnon energies  of
MnSi and MnC for two selected directions in the reciprocal space.
We see that the MnC curves are flat in the second half of the
\textbf{q} interval showing here a very weak dispersion. On the
other hand, the MnSi curves are much closer to a simple
cosinusoidal form with considerable dispersion in this part of the
\textbf{q} interval. Comparing now the low-\textbf{q} parts of the
curves we notice that MnC curves lie higher than the MnSi curves.
Therefore, the relative contribution of the low-energy magnons to
the RPA value of the Curie temperature is less important in MnC
compared to MnSi.

This combination of features of the wave-vector dependencies of
the frozen-magnon energies is responsible for a much smaller
difference between the RPA and MFA estimations of the Curie
temperature in the two cases. Since the form of the magnon
dispersion is reflected on the properties of the inter-atomic
exchange interactions it is instructive to compare the patterns of
the exchange interactions of the two systems. Taking into account
that only Mn-Mn interactions contribute importantly in the Curie
temperature we will focus on the comparison of these interactions.

The analysis of the Mn-Mn exchange interactions in MnSi and MnC
(figure~\ref{fig4}) shows a strong difference between the two
systems. In MnSi, only the first-neighbor exchange interaction
plays important role. All other interactions are much smaller. On
the other hand, in the case of MnC the second-neighbor interaction
is comparable with the first-neighbor one. The further few
interactions though sizable play less important role. The presence
of two first coordination spheres with similar values of the
exchange interactions, on the one hand, makes the exchange
interaction in MnC more long ranged and, on the other hand, it
effectively increases the coordination number of the atoms with
leading exchange interaction. Either of these properties leads to
a better accuracy of the MFA estimation.

\section{Summary and Conclusions}\label{sec6}

We have presented a detailed study of the stability of the
ferromagnetism in six transition-metal compounds crystallizing in
the zinc-blende structure. We calculated from first-principles the
electronic structure and employed the frozen-magnon approximation
to evaluate the exchange parameters and the Curie temperature. In
the calculation of the Curie temperature we used both the
mean-field and random-phase approximations.

Considering the CrAs and CrSe compounds we found in the case of
CrAs a very high Curie temperature exceeding 1000 K. The
ferromagnetism of CrAs is stable with respect to the compression
of the lattice. On the other hand CrSe possesses about two times
lower Curie temperature which decreases strongly with lattice
compression.

The properties of MnSi and MnGe are similar to the properties of
the isovalent CrSe while the properties of MnAs are similar to the
properties of isovalent CrSe.

MnC presents an interesting case that differs from all other
systems studied. This system is also half-metallic with the
half-metallic gap in the majority-spin channel in contrast to
other systems where the half-metallic gap is in the minority-spin
subsystem. The Curie temperature of MnC was found to be about 500
K.

In all cases considered we found that half-metallicity is
favorable for ferromagnetism and leads to an increased Curie
temperature.

We conclude that two compounds, MnSi and CrAs, possess properties
which make them promising candidates for spintronics applications:
large half-metallic gap and high Curie temperature. The stability
of the Curie temperature in these systems allows to expect that
being grown on the substrate of a binary semiconductor they will
preserve the desired magnetic properties.

\ack The financial support of Bundesministerium f\"ur Bildung und
Forschung is acknowledged.

\end{document}